# Characterization of cortical motor function and imagery-related cortical activity: Potential application for prehabilitation

M. Korostenskaja*, C. Kapeller, K.H. Lee, C. Guger, J. Baumgartner, E.M. Castillo

M. Korostenskaja
Functional Brain Mapping and BCI Lab
Florida Hospital for Children
Orlando, USA
*milena.korostenskaja@gmail.com

M. Korostenskaja, E.M. Castillo
MEG Center
Florida Hospital for Children
Orlando, USA

M. Korostenskaja, Ki H Lee, J. Baumgartner, E.M. Castillo
Florida Epilepsy Center
Florida Hospital
Orlando, USA

C. Kapeller, C. Guger
g.tec Guger Technologies OG
Schiedlberg, Austria

*Abstract*—To minimize functional morbidity associated with brain surgery, new preventive approaches (also referred to as "prehabilitation") by using motor-imagery-based computer interfaces (MI-BCIs) can be utilized. To achieve successful MI-BCI performance for prehabilitation purposes, the characteristics of an electrocorticographic (ECoG) signal that is associated with overt motor function ("real movement" - RM) versus covert motor function ("motor imagery" - MI) need to be determined. In our current study, 5 patients with pharmacoresistant epilepsy (2 males, average age 25 years, SD 15), undergoing evaluation for epilepsy surgery participated in both RM and MI tasks. Although the RM- and MI- related ECoG changes had some common features, they also differed in a number of ways, such as location, frequency ranges, signal synchronization and desynchronization. These similarities and differences are discussed in a view of other neuroimaging studies, including magnetoencephalography (MEG) and functional magnetic resonance imaging (fMRI). We emphasize the need for inclusion of a broad spectrum of frequencies in ECoG analysis, when RM- and MI- related activities are concerned.

*Keywords*—brain-computer interfaces (BCIs); brain surgery; electrocorticography (ECoG); epilepsy surgery; functional mapping; motor function; motor imagery; motor-imagery-based BCIs (MI-BCIs); motor rehabilitation; prehabilitation; tumor surgery

## I. Introduction

Brain surgery for tumor removal or for eliminating seizure onset zone are often the only viable treatment option to improve the quality of life of people suffering from neurological disorders, such as brain tumors or pharmacoresistant epilepsy, respectively [1]. However, in some cases the surgical target zone overlaps with eloquent (functionally significant) cortex, severe post-surgical deficits can occur (for example, hemiplegia exacerbated by functional hemispherectomy) [2]. To restore the impaired function (for example, hand movement) an intensive and prolonged in time (from several month to several years) rehabilitation process is required [3]. Moreover, some of the functions can be never completely restored. Indeed, potentially curative surgical procedures are often postponed or not considered simply due to the possible anticipated functional impairments. For a number of brain tumor surgeries even partial overlap between surgical resection zone and eloquent cortex may result in a reduced resection margin, thus compromising the effects of the resection on post-surgical tumor development prognosis.

We propose a new preventive approach with the goal of minimizing the functional morbidity associated to brain surgery. This approach involves a presurgical rehabilitation of those functions at risk, facilitating mechanisms of an interhemispheric or intrahemispheric transfer of potentially compromised functions. In other words, instead of limiting patients to post-surgical rehabilitation only, a pre-surgical rehabilitation can be used to minimize post-surgical deficits and the time spent in rehabilitation facilities. For some patients, this approach could allow surgical tumor/epilepsy treatment option that was not possible without it. The feasibility of such approach has been demonstrated in the most recent work by Rivera-Rivera et al. [4], where they have introduced a term "prehabilitation" as a substitute for "presurgical rehabilitation". They have applied a prehabilitation treatment in a form of cortical stimulation to 5 patients scheduled for brain glioma surgery, which resulted in a possibility to expand tumor resection margin, thus leading to better post-surgical outcomes.

In our opinion, motor imagery-based brain-computer interface (MI-BCI) has a strong potential to become a successful tool for prehabilitation of motor function. Although this tool is relatively new in the world of motor rehabilitation, it has already demonstrated high impact on rehabilitation outcomes, for example in stroke patients [5-8]. MI-BCI is a technique that records signals produced by patient's brain during motor imagery of the impaired limb and translates these signals into commands that can trigger the movement of a virtual limb (i.e., presented on a pc screen), a prosthetic limb or/and the electrical stimulation of muscle groups to facilitate the movement of the spastic limb. For the successful work of MI-BCIs, it is very important, therefore, to be able to extract the correct features related to imaginary motor activity, such as spatial (location), temporal (time), and spectral (frequency) signal characteristics that can be effectively further translated into computer signals. A number of studies demonstrate the similarities between signals related to real motor movement and the imaginary one [9], whereas other studies show that these two signals can differ in their characteristics [10]. Therefore, comparative studies must be conducted to clarify this question.

One of the signals that can be used for the purpose of MI-BCI pre-habilitation is the one recorded by means of electrocorticography (ECoG). This signal is currently utilized in a real-time fashion for a novel functional mapping paradigm in patients that are being evaluated for epilepsy surgery or undergoing tumor surgery. Introduced by Schalk et al. [11] and validated by our team in both pediatric patients [12] and adults [13], real-time functional mapping (RTFM) with ECoG represents a viable alternative to current invasive functional mapping approaches for localization of the eloquent cortex, such as electrical cortical stimulation mapping, and contributes towards improvement of post-surgical functional outcomes [14]. High accuracy results were achieved by using both in-house built [15] and commercial [16, 17] RTFM systems. The ECoG recorded and processed in a real time can become a useful tool for pre-habilitation purposes in brain tumor and epilepsy surgery patients. Therefore, in this preliminary study, we aimed to evaluate similarities and differences between the ECoG signal recorded in response to a real motor movement (real movement - RM) and the imaginary one (motor imagery - MI). The implications of our results towards maximizing MI-BCI performance for prehabilitation purposes are discussed. The comparison of our results with the results from non-invasive functional mapping modalities is presented.

## II. METHODS

### A. Study Participants

The ECoG was recorded from 5 patients with pharmacoresistant epilepsy (2 males and 3 females, average age 25 years, SD 15) (TABLE I), undergoing evaluation for epilepsy surgery with implanted ECoG electrodes. Patients gave their informed consent to participate. The study was approved by the institutional review board (IRB) of Florida Hospital. All of the subjects had a single hemisphere coverage (3 of them - left and 1 of them - right). One of the subjects, on the other hand, had a very unique electrode coverage, implemented only in a very

TABLE I. SUBJECT DESCRIPTION

| Subject's ID | Age, years | Gender[a] | Dexterity[b] | Development, range | Grid placement, hemisphere[b] | Number of sampled electrodes |
|---|---|---|---|---|---|---|
| S27 | 12 | F | R | high average | R | 128 |
| S30 | 12 | M | L | mild cognitive delay | L | 128 |
| S31 | 20 | F | R | average | L | 128 |
| S32 | 32 | M | R | low average | L&R | 128 |
| S33 | 47 | F | R | high average | L | 128 |

[a.] F - female, M - male; [b.] R - right, L - left

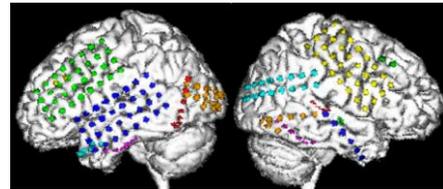

Fig. 1. Bilateral hemispheric coverage with ECoG electrodes (strips and grids) in study participant #32. The coverage extended to left frontal, parietal and occipital lobes, as well as to the right fronto-central, temporal and occipital areas.

few surgical centers worldwide. It was extending to both cerebral hemispheres and its placement was achieved by means of bilateral craniotomy (Fig. 1). All grids were implanted exclusively for surgical consideration.

### B. Experimental Design

The recording of ECoG activity was performed during RM and MI tasks requiring from patients to open and close their hand or imagining doing it, respectively. The hand chosen to perform the movement was the one contralateral to the hemisphere with ECoG coverage. In case of bilateral coverage for participant #32, both hands (L and R) were performing the task (one at a time) and corresponding activity was recorded from both hemispheres. The data have been recorded in the manner of block design [18]. First, 6 minutes of resting activity was recorded followed by 30 s blocks of control task and task-related activity (Fig. 2). Study participants had possibility to do a 5 minute training before the actual recording procedure. The subjects were asked to remain silent during the task execution and follow the instructions presented on a screen that was placed in front of them.

### C. Data Acquisition

The ECoG data were recorded at patients' bedside by using a set of biosignal amplifiers (g.tec medical engineering GmbH, Schiedlberg, Austria) during patients' stay in the hospital for pre-surgical evaluation for epilepsy surgery. A sampling rate

1200 Hz was utilized. A built-in band-pass filter was set to record a range of frequencies within 0.5 - 500 Hz. A BCI2000 software [19] was utilized for task presentation and as an interface between the task-generating computer and amplifier.

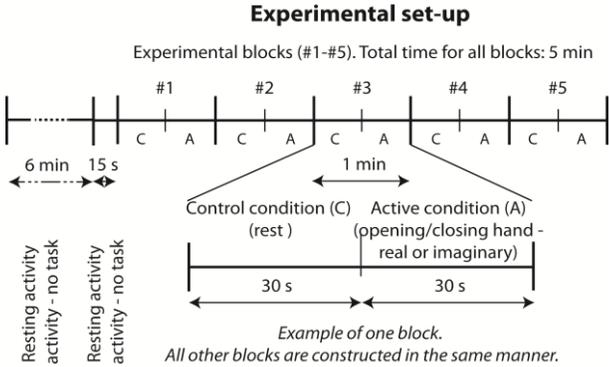

Fig. 2. Experimental set-up.

*D. Data Analysis*

Data were reviewed for artifacts and noisy channels in BCI2000 Viewer software [19]. Noisy channels were excluded from further analysis, which has been performed with CortiQ software package [16, 17]. The data were filtered with notch filters at all harmonics *hi* of the 60 Hz power line frequency (a notch width of *hi* ± 5 Hz). We have included the following frequency ranges of interest for our analysis of ECoG data: 8 - 13 Hz, 8 - 28 Hz, 18 - 28 Hz, 18 - 28 Hz, 30 - 50 Hz, 70 - 170 Hz, 250 - 350 Hz. In the μ- and β- frequency ranges we expected to detect event-related desynchronization (ERD) with respect to motor tasks. The combined μ- /β- frequency bands were chosen, as they can provide better features selection for BCI control.

For all of them a corresponding band-pass filter (Butterworth filter of order 5) was used. The main areas of comparison between RM and MI conditions were the following: Signal elicitation, localization and distribution, as well as spectral signal characteristics and synchronization/desynchronization with the task.

III. RESULTS

*A. Signal Elicitation, Localization and Distribution*

All main results are summarized in TABLE II. Four (80%) out of five study participants demonstrated significant responses during the RM task and only three (60%) - during the MI task. Significant changes ECoG in signals for RM task were localized in the vicinity of sensory-motor cortex for all four study participant (100%), whereas not all significant responses elicited during the RM task had similar localization. However, the areas of activation did not completely overlap between the real and imaginary movement conditions. Importantly, in both conditions some additional activation was found in auditory cortex - as a response to some audio-prompts provided by the researcher conducting the study. In addition, the significant areas of activation during MI task were less distributed than during the RM task and had fewer significantly activated electrode sites. Interestingly that motor imagery of one hand was associated with the significant changes in ECoG activity in both hemispheres (Fig. 3).

*B. Spectral Signal Characteristics and Synchronization Properties*

In regards to the frequency range (see, TABLE II), where the significant activations during the task execution were observed, the frequency spectrum for RM was very broad and encompassed frequencies from alpha (8 Hz) to high gamma (170 Hz) (Fig. 3). At the same time, the frequency spectrum corresponding to significant changes of ECoG signal during MI task execution tended to be in a higher range of frequency spectrum, such as beta (18 Hz) to high gamma (170 Hz) (Fig. 3). Moreover, there were differences between signal synchronization/desynchronization patterns between RM and MI conditions. The RM condition was characterized by both signal synchronization and desynchronization during the task execution, with synchronization corresponding to a high gamma frequency range (70-170 Hz). Other observed synchronization during the RM task in 8-13 Hz and 2-28 Hz frequency range was localized in auditory cortex and might be associated with auditory prompts - not the motor task execution. At the same time, all our recorded ECoG signal associated with imagery task was characterized by desynchronization during the task execution.

IV. DISCUSSION

In our current study, we have compared the characteristics of ECoG signals recorded during the real-movement and motor imagery tasks with the aim to benefit future signal processing and feature extraction for motor-imagery-based BCI application for motor rehabilitation purposes, and, specifically for "in advance" rehabilitation - also referred to as "prehabilitation". Both commonalities and differences in the ECoG signals between these two conditions were observed.

Whereas significant activation during RM condition was observed in 83% of study participants, only 60% of participants elicited significant MI-related response. This finding is in line with previously published data regarding the MI-BCI adoption rate: according to Jeunet et al. [20], only 70-90% of users are able to achieve MI-BCI "literacy" (ability to successfully control BCI by means of motor imagery). However, consideration of both external and internal factors, affecting MI-BCI performance, such as training [21], feedback [22], neurophysiological signal analysis [23], system's adaptability [24] and other factors might contribute towards the significant increase of MI-BCI "adoption" rates [20]. Moreover, in our current study we have demonstrated that MI-related activity varies on individual basis, which implies that individual variability must be taken into consideration when working with MI-BCIs. It is in line with previously conducted studies [25], where subject-specific models used for MI-BCIs control can more accurately interpret users' neurophysiological signals [23].

ECoG findings demonstrated that the cortical MI-related activations differed from those related to RM. Although hand motor activation during RM was clearly localized in the sensory motor cortex 4 patients, the MI-associated cortical activation was not universal. Although it resembled the RM-related activation to certain extent, it was not identical. Some non-invasive neuroimaging (for example, fMRI) studies have indicated no or very subtle activation of sensory-motor cortex during motor imagery when compared to real-movement-related activation [26]. However, our study is partially in line with other recent findings from magnetoencephalography (MEG) [9] and fMRI [10] indicating that MI-related activation takes place, although it differs in some aspects from RM-related cortical activation.

TABLE II. COMPARISON BETWEEN ECoG ACTIVITY RECORDED DURING THE REAL MOTOR MOVEMENT AND MOTOR IMAGERY TASKS

| | Real Motor Movement (RM) Task | Motor Imagery (MI) Task |
|---|---|---|
| significant signal elicitation | 80% (4/5) | 60% (3/5) |
| significant signal localization | Vicinity of sensory-motor cortex | Vicinity of sensory-motor cortex, Frontal cortex |
| significant signal frequency range, Hz | S27: 8-28<br>S30: -<br>S31: 8-28, 18-28, 30-50, 70-170;<br>S32 (Right hand movement, Left hemisphere): 8-28, 18-28, 70-170;<br>S32 (Left hand movement, Right hemisphere): 8-13, 8-28, 18-28, 70-170;<br>S33: 8-13, 8-28. | S27: 18-28, 30-50;<br>S30: - ;<br>S31: - ;<br>S32 (Right hand movement imagination, Left hemisphere): - ;<br>S32 (Right hand movement imagination, Left hemisphere): 70-170;<br>S33: 18-28. |
| significant signal synchronization vs desynchronization with the task | S27: desynchronization in 8-28;<br>S30: - ;<br>S31: desynchronization in 8-28 Hz, 18-28 Hz, 30-50 Hz and *synchronization in 70-170 Hz*;<br>S32 (Right hand movement, Left hemisphere): both synchronization and desynchronization in 8-13 Hz and 8-28 Hz, desynchronization in 18-28 Hz, and *synchronization in 70-170Hz*;<br>S32 (Left hand movement, Right hemisphere): desynchronization in 8-28 Hz, 18-28 Hz; *synchronization in 70-170Hz*;<br>S33: desynchronization in 8-13 Hz, 8-28 Hz. | S27: desynchronization in 18-28, 30-50<br>S30: - ;<br>S31: - ;<br>S32 (Right hand movement, Left hemisphere): *desynchronization in 70-170Hz*;<br>S32 (Left hand movement, Right hemisphere): -<br>S33: desynchronization in 18-28 Hz. |

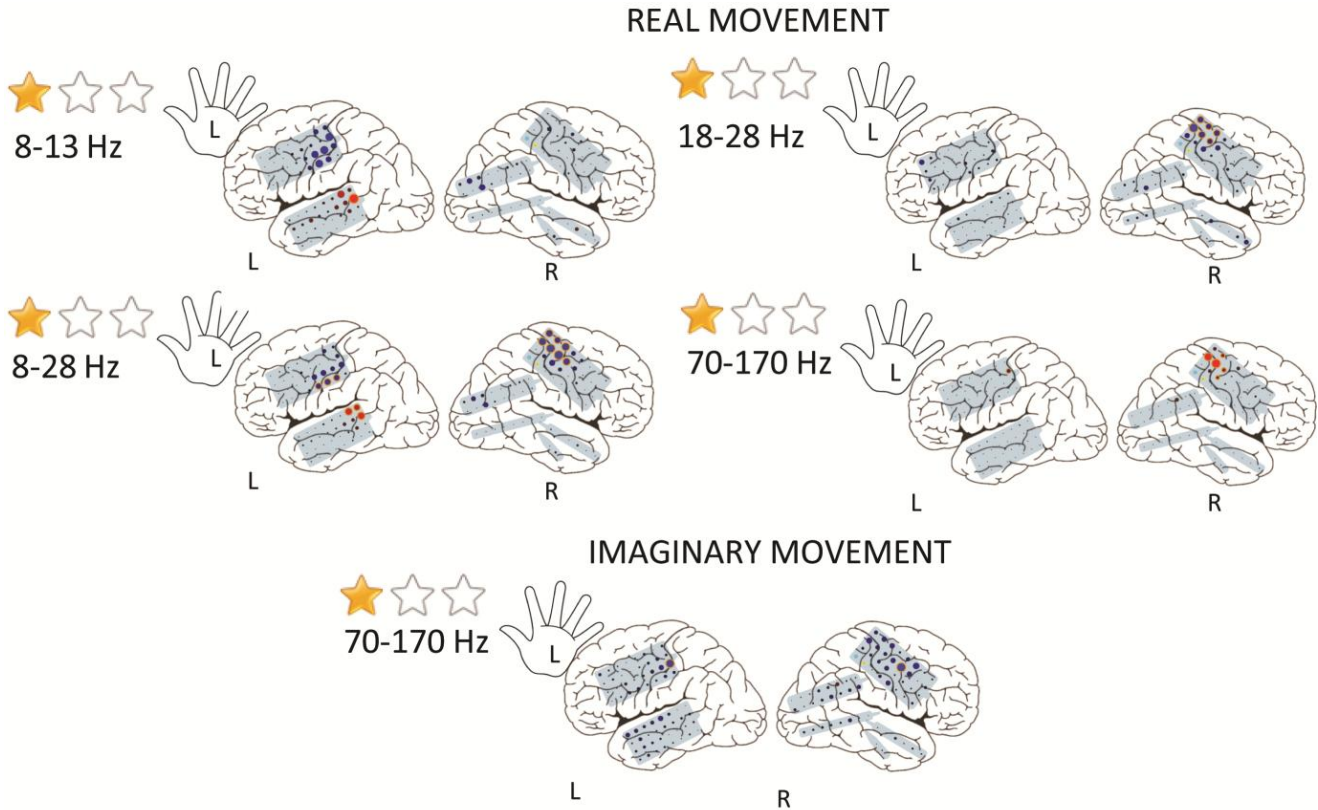

Fig. 3. Real and imaginary hand movement ECoG results for study participant #32. The significant values are indicated in circled red/blue blobs. Desynchronization of the signal during the task execution is presented in blue and synchronization in red. Please, note activation that is not related to motor activity in temporal lobes.

Although gamma activity has been proposed as the main frequency range allowing to reliably perform functional mapping of eloquent cortex [27], our study demonstrates that other frequency bands can also provide significant results and should be taken into consideration. Importantly, particular attention must be paid to a lower than gamma frequency ranges when motor function and motor imagery is concerned. This is in line with the previous study by Kapeller et al [28], demonstrating that by considering a wide spectrum of frequencies and particularly including lower ones, a high performance (95.4 % accuracy) of MI-BCI controlled tele-operation of a humanoid robot was achieved. Our results are also in line with a RM and MI ECoG signal comparison study by Miller et al. [29]. It should be noted that in this study we have not found any significant task-related activation in ECoG signal in a higher than 170 frequency range (specifically 170-250 Hz), which demonstrates the upper cut-off limit of the useful signal frequency range analysis at approximately 170 Hz.

A level of cognitive development of a patient needs to be taken into consideration when working with RM and MI tasks, as in our study the only person, who produced the data that did not have significant activation during either RM or MI tasks, was the one with mild cognitive delay (assessed by means of neuropsychological evaluation, see Table 1). Specific strategies need to be considered to improve task performance in patients with this level of cognitive development.

Our ultimate goal is to facilitate the mechanisms of an interhemispheric or intrahemispheric transfer of potentially compromised motor function due to the brain surgery. Our results (Fig. 3) demonstrate that when ECoG is utilized for data acquisition, it is possible to observe both ipsilateral and contralateral movement-related and imagery- related activity when only one hand is used for task execution. This observation is very promising for prehabilitation approaches involving intrahemispheric transfer of motor function.

V. LIMITATIONS, CONCLUSIONS AND FUTURE PERSPECTIVES

*A. Study Limitations*

We have detected some significant non-motor-related activation during both RM and MI tasks primarily in the auditory cortex, which, in our opinion, was associated with the study participant listening to the prompts provided by the study conducting research scientist. In this case, to eliminate such contamination on signal analysis and classification, the analysis of the ECoG activity might be restricted to the sensory-motor cortex only, like for example, in an ECoG study by Kapeller at al. [30] and MEG study by Sugata et al. [9].

Of course, our current study was limited by its sample size and is preliminary in nature. Larger scale studies are warranted to expand upon these preliminary findings.

*B. Conclusions*

The conclusions of our study are as follows:

1. MI-related activity can be detected similarly to RM-related activity, however, in a smaller number of people. The cognitive status of a study participant may influence both MI- and RM- related ECoG mapping results;

2. Compared with RM-, the MI-based activation is less universal and may vary on individual basis. The assessment of individual MI-related cortical activation patterns and utilizing them to improve MI-BCI performance are warranted. The utilization of signals from MI-related substrates different than M1 can be of high importance for MI-BCI-based rehabilitation of patients with damaged/affected M1.

3. The analysis of both RM- and MI-related activity should incorporate all available frequency ranges. At the same time, an upper cut-off limit for signal analysis might be considered at approximately 170 Hz.

4. Procedures helping to avoid the contamination of the analysis of MI- related activity by the activity related to auditory or visual information processing should be in place to detect significant MI-BCI signals utilized for pre-habilitation.

5. Both contralateral and ipsilateral real motor- and motor imagery- related activity can be observed when only one hand is used for task execution. This can have a direct impact on a possibility to facilitate intrahemispheric transfer of potentially compromised motor function ("prehabilitation") due to brain surgery.

*C. Future Perspectives*

Combining information from invasive (ECoG) and non-invasive (MEG) imaging modalities should be considered aiming at providing valuable insight into the neural origins of RM and MI as well as their application for pre-habilitation. This is particularly important, as both invasive (ECoG-based) and non-invasive (EEG- or MEG-based) pre-habilitation options should be considered.

We also propose implementation of machine learning algorithms in the analysis of relevant ECoG functional mapping data and its application for prehabilitation.

ACKNOWLEDGMENT

We would like to thank Drs. G. Schalk and P. Brunner for providing us with their in-house built version of BCI200-based software and for their continued support for our ECoG-related studies.